%% file: main.tex
\documentclass[a4paper, aps, superscriptaddress, amsfonts, amssymb, amsmath, reprint, showkeys, nofootinbib]{revtex4-2}
\usepackage[english]{babel}
\usepackage[utf8]{inputenc}
\usepackage[colorinlistoftodos, color=green!40, prependcaption]{todonotes}
\input{preamble}
\graphicspath{{Figures/}}
\begin{document}

\title{FIR transmission and resistively detected cyclotron resonances in InSb}

\author{M.E. Bal}
\email[Correspondence email address: ]{maurice.bal@ru.nl}
\author{N. De{\ss}mann}
\author{K. Saeedi}
\author{U. Zeitler}
\email[Correspondence email address: ]{uli.zeitler@ru.nl}
\affiliation{HFML-FELIX, Toernooiveld 7, 6525 ED Nijmegen, The Netherlands}
\affiliation{Institute for Molecules and Materials, Radboud University,Heyendaalseweg 135, 6525 AJ Nijmegen, The Netherlands}

\date{\today} 

\begin{abstract}
We have developed an experimental setup to simultaneously acquire magneto-transmission spectra and measure the photoconductive response. The low-temperature ($T$=4.2 K) magneto-transmission data for weakly doped InSb in the frequency range 3-15 THz and magnetic fields up to 25 T, shows that the AC conductivity is governed by the classical Drude response. Moreover, the resulting light-induced voltage in this system, consisting of thermoelectric and resistive contributions, both contain features that are related to cyclotron resonances. The frequency dependence of this resonance is in good agreement with $\mathbf{k\cdot p}$ perturbation theory and corresponds to transitions between the first and second Landau level with spin-up electrons. The obtained effective mass $m^{*}(B=0)=0.014m_{e}$, is in line with the value extracted from magneto-transmission experiments. Additionally, we can observe evidence of a fluence-dependent metal-insulator transition in the thermoelectric signal, which suggest that the electronic system becomes less sensitive to the light at higher frequencies. 
\end{abstract}

\maketitle

\section{Introduction}
In order to satisfy the demand for faster electronics, eventually making the transition from GHz operating frequencies to those in the THz regime, it has become crucial to investigate and understand the electronic properties of semiconductors in this frequency range. Due to the limitations of electronic measurement equipment, one cannot straightforwardly probe these electronically. One solution is to probe these properties optically by measuring the magneto-transmission, which contains information about the AC conductivity. Besides for investigating the AC conductivity, infrared (IR) spectroscopy is also a powerful tool for probing the band parameters of semiconductors. In particular, THz radiation can be used to induce various excitations, ranging from cyclotron \cite{PhysRev.122.475,PhysRevB.105.205204}, impurity cyclotron \cite{PhysRevB.31.3560,PhysRevLett.40.1151}, spin \cite{Dobrowoska_1990,PhysRev.181.1206} as well as phonon-assisted resonances \cite{PhysRev.182.790}, which give valuable information about the effective mass, impurity states, effective $g$-factor and phonon modes, respectively. Experimentally, these excitations are usually detected as a dip in the magneto-transmission signal. However, this limits its application to devices with dimensions similar to the spot size of the incident light. Performing these experiments on systems such as nanowires is therefore not possible as the absorption feature is not discernible above the background signal. Another possibility to detect these resonances, which is compatible with nano-devices, would be to measure the photoconductive response of the system \cite{mani2012observation,PhysRevB.76.081406,stachel2014cyclotron}. Irradiating the sample with infrared radiation will cause an increase in temperature and change its electronic properties. This absorption will be largest at resonance, resulting in a strong modulation of the photoconductive response.

In this paper, we present far-infrared (FIR) transmission data and our proof-of-principle experiments for resistively detecting CRs in weakly doped InSb. Firstly, the frequency dependence of conductivity is investigated by comparing the magneto-transmission spectra to calculated spectra based on the DC conductivity. The simultaneously acquired light-induced electrical signal, which mainly has a thermoelectric contribution, displays a strong response at resonance. The position of these cyclotron resonances is consistent with those observed in the transmission data. Similar features are seen in the resistive photoconductive response, though the signal strength is can be up to two orders of magnitude smaller.

\section{Samples and Methods}
\begin{figure}[t]
    \centering
    \includegraphics[width=8.4cm]{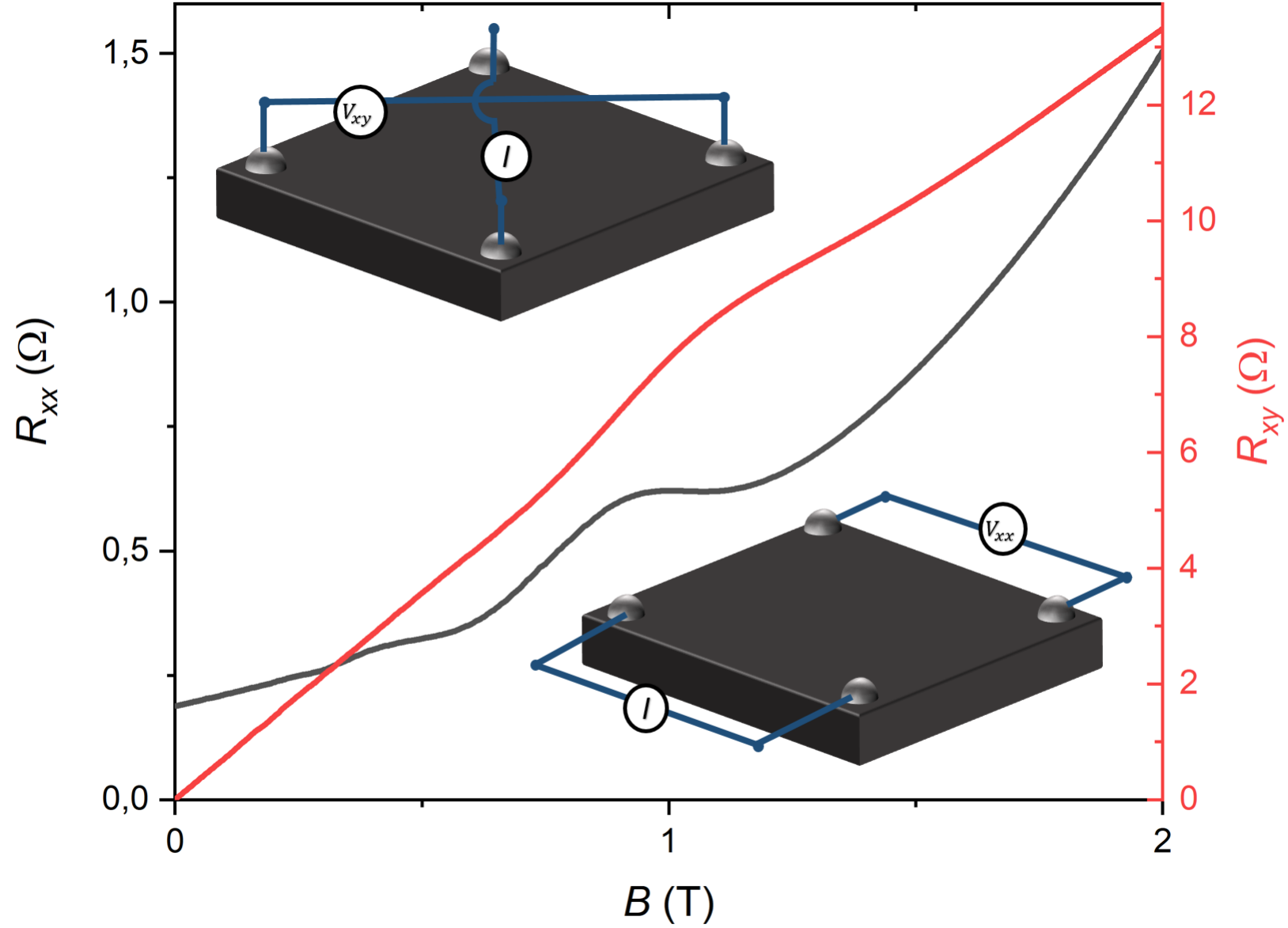}
    \caption{Longitudinal and transverse resistance as a function of magnetic field for weakly doped InSb at 4.2 K. The insets show the contact configuration to measure the individual resistance components.}
    \label{fig:1}
\end{figure}

The weakly doped InSb samples discussed in this paper, were cut from a 500 $\mu$m-thick commercially purchased 2" wafer supplied by SurfaceNet GmbH, grown using the Czochralski method. The resulting 4$\times$4 mm$^{2}$ crystals of nominally undoped (100) $n$-InSb were electrically contacted by fabricating In contacts that were arranged in the van der Pauw geometry. Standard low-frequency lock-in techniques were used to acquire $R_{xx}$ and $R_{xy}$ (see inset Fig. \ref{fig:1}). Using the classical Hall effect, we extract that our InSb crystal has a carrier concentration of $n$=1.41$\times$10$^{15}$ cm$^{-3}$ at 4.2 K with a corresponding mobility of $\mu$=8.32$\times10^{4}$ cm$^{2}$V$^{-1}$s$^{-1}$.

At HFML-FELIX there is the possibility to couple tunable intense IR radiation from the free electron laser (FEL) into a 33 T Florida-Bitter magnet, enabling us to perform spectroscopy measurements at high magnetic fields \cite{perenboom2013}. The pulsed THz radiation is guided from the optical cavity to the magnet via a 90 m long evacuated beamline, where it is focused into a cylindrical, oversized, brass waveguide and passed through a diamond window, into a new-silver waveguide that was inside a $^4$He bath cryostat. In order to perform magneto-transmission experiments, THz radiation must be able to pass through the cryostat. Hence, the cryostat was equipped with two diamond windows, one between the He bath and the vacuum space and the other between the vacuum space and the ambient air. Brass cones, which focus the THz radiation, are placed before and after the sample that is mounted in the Faraday configuration. The sample is placed in a leadless chip carrier (LCC) with a hole drilled into the center, which allows the transmitted light to propagate through the sample space. A copper foil mask was placed on top of the LCC to prevent unwanted illumination of the contacts that will result in rectification.

Another brass waveguide, which is evacuated with a large scroll pump, is used to guide the transmitted THz light from the bottom of the cryostat toward an IR photo-detector. Depending on the frequency, we either use an InSb (0.3-3 THz) or Ge:Ga (2.5-30 THz) detector that are mounted in a QMC or IR labs cryostat, respectively. These detector elements are cooled down to liquid He temperatures (4.2 K) to increases the sensitivity of the InSb hot electron bolometer and to induce carrier freeze-out in the Ge:Ga bolometer. Optical density filters, suitable for THz frequencies, are used as attenuators in case the incoming IR radiation saturates the detector. The transmittance $T$ is obtained from the photo-detector signal $V_{bolo}$ by using the following definition

\begin{equation}
    T(B,\omega)=\frac{V_{bolo}(B,\omega)}{\mathrm{max}(V_{bolo}(B,\omega))}.
\end{equation}

The photoconductive response of the sample was recorded in order to resistively detect cyclotron resonances. For this purpose, the sample was biased with current pulses that are generated using a SRS DG645 digital delay generator in combination with a Howland circuit. A voltage follower subcircuit was included to ensure that the internal resistors in the AMP03 unity gain amplifier are still balanced regardless of the value of the shunt resistor $R_{s}$, i.e. decoupling the shunt resistor from the rest of the Howland circuit. The timing of the pulse train was such that the initial pulse coincides with the FEL macropulse, whereas the second one arrives after the light pulse. The light-induced voltage changes $\Delta V=V_{light}-V_{dark}$, which correspond to the difference between the two pulses, were amplified using a NF SA-430F5 (46 dB gain) in series with a FEMTO DHPVA-101 (10-60 dB gain) amplifier, increasing the signal amplitude to hundreds of mV, which was well within the dynamic range of the NI PXIe-5162 oscilloscope that was used for data acquisition. A measurement program was developed that could directly perform background subtraction by employing a boxcar-like acquisition scheme. By simultaneously measuring the signal from the detector, we were able to correlate cyclotron resonance signatures in the optical signal to those in the electrical signal. At this point, it should be noted that for these measurements we used an alternative method to couple the light into the above mentioned setup. A free space optics configuration was utilized to collimate the light before directing it towards the new-silver waveguide with gold-plated copper mirrors. In order to minimize light absorption, the entire optical path is situated in a nitrogen purge box. The biggest advantage of this method is that one has much more control over the optical path of the light in the measurement setup, compared to the waveguide where one can only tweak the last two mirrors of the beamline. This allows us to reliably place the laser spot in the center of the sample. Although, the overall laser power that reaches the sample will be lower, as a substantial part of the light is absorbed by the lenses.  

\section{Results}

\begin{figure*}[t]
    \centering
    \includegraphics[width=16cm]{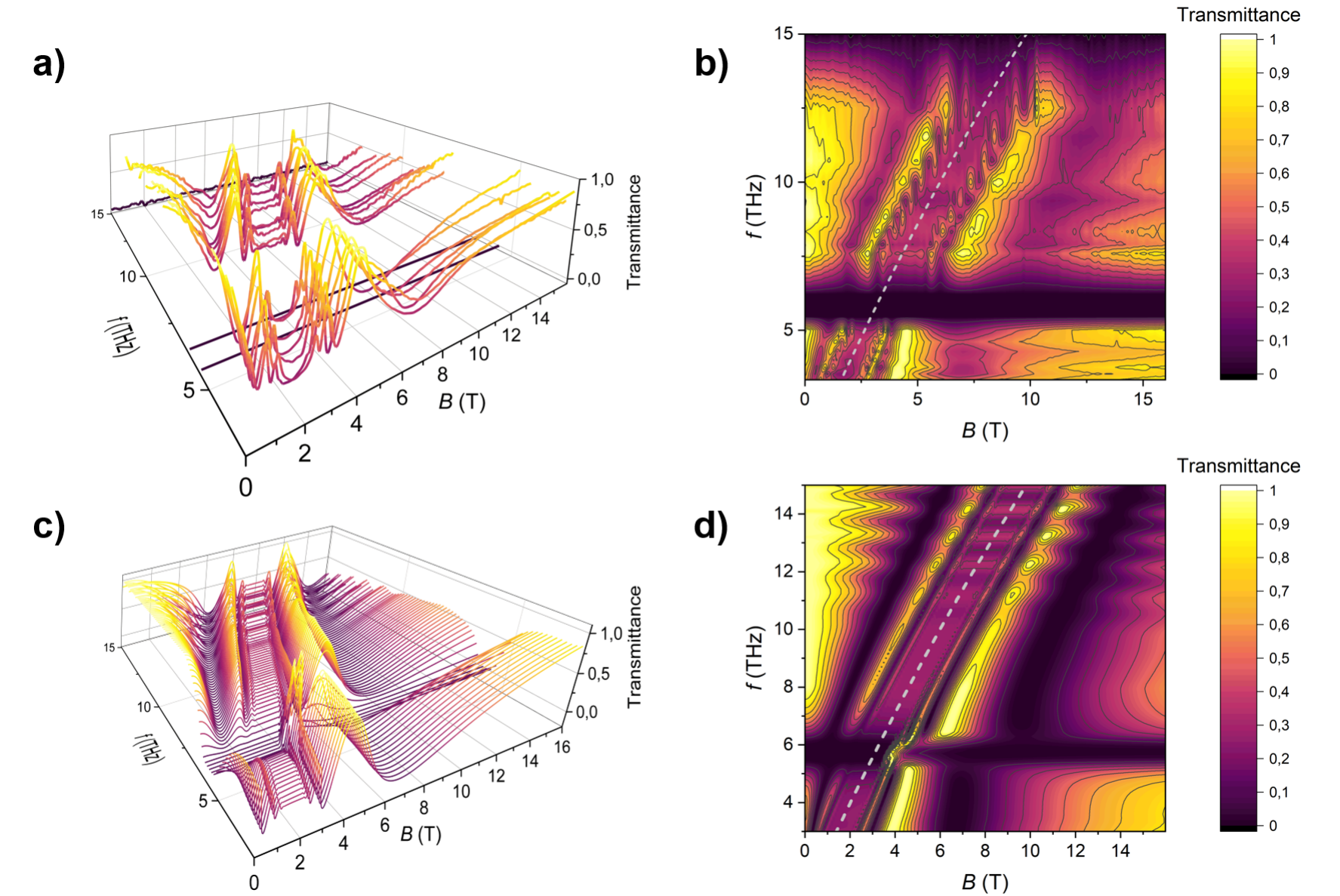}
    \caption{Waterfall (a) and contour (b) plot of magneto-transmission spectra that were measured at different frequencies in a slab of weakly doped InSb at $T=4.2$ K. Across the entire frequency range the power of the incident radiation varied between 0.01-0.5 mW. By assuming that the conductivity is described by the Drude model we can simulate the corresponding magneto-transmission spectra, which are displayed in a waterfall (c) and contour (d) plot. For the calculations we used a carrier concentration of $n=1.41\times10^{15}$ cm$^{-3}$ and a scattering rate of $\nu=1$ THz. The dashed grey line indicates the theoretically predicted position of the CR.}
    \label{fig:2}
\end{figure*}

\subsection{FIR transmission}
The magneto-transmission spectra of our InSb sample are displayed in Fig. \ref{fig:2}a and b for magnetic fields up to 16 T in the frequency range 3-15 THz. In these magneto-transmission spectra we can identify a broad cyclotron resonances surrounded by interference fringes as well as two frequency regions where the sample is completely opaque. The first region between 5 and 6 THz corresponds to the Reststrahlen band of the transverse (TO) and longitudinal (LO) optical phonon mode, which have a natural frequency of 5.4 and 5.85 THz, respectively \cite{PhysRevB.12.2346}. The opacity for radiation with frequencies larger than 15 THz, most likely stems from the 2LO phonon mode ($f_{2LO}$=12.35 THz \cite{koteles1974far}). Similar magneto-transmission spectra were observed by Wang \textit{et al.} who performed time-domain THz magneto-spectroscopy measurements on InSb \cite{Wang2010}. 

Using the theoretical model described in their work, we can try to reproduce our experimental results and investigate whether it still holds in the frequencies range far beyond the scattering rate. We start by introducing an expression for the dielectric tensor $\epsilon$, which determines the optical properties of materials \cite{Palik_1970}

\begin{equation}\label{eq2}
    n^{2}=\vec{\epsilon}(B,\omega)=\epsilon_{r}+\epsilon_{ph}+\frac{i\vec{\sigma}(B,\omega)}{\omega\epsilon_{0}},
\end{equation}

\noindent where $\epsilon_{0}$ is the permittivity of free space, $\epsilon_{r}$ is the dielectric permittivity, $\omega$ is the angular frequency, $\vec{\sigma}(B,\omega)$ is the complex conductivity tensor and $\epsilon_{ph}$ accounts for the electric dipole moment generated by optical lattice vibrations in polar semiconductors such as InSb. The contribution of optical lattice vibrations is included by employing the harmonic oscillator approximation, i.e. $\epsilon_{ph}=\epsilon_{r}(\omega_{L}^{2}-\omega_{T}^{2})/(\omega_{T}^{2}-\omega^{2}-i\gamma_{ph}\omega)$ \cite{Palik_1970}. $\omega_{L}$, $\omega_{T}$ and $\gamma_{ph}$ are the angular frequencies of the LO, TO phonons and the phonon scattering rate, respectively. In the Faraday configuration Eq. \ref{eq2} can be rewritten as 

\begin{equation}\label{eq3}
    \begin{split}
    n_{\pm}^{2}&=\epsilon_{\pm}(B,\omega)=\\
    &\epsilon_{r}\left(1-\frac{\sigma_{xx}\mp i\sigma_{xy}}{i\omega\epsilon_{0}}+\frac{\omega_{L}^{2}-\omega_{T}^{2}}{\omega_{T}^{2}-\omega^{2}-i\gamma_{ph}\omega}\right),
    \end{split}
\end{equation}

\noindent here $\sigma_{xx}$ and $\sigma_{xy}$ are the real longitudinal and transverse components of the conductivity, respectively. The left-circularly polarized (-) mode corresponds to the motion of a negatively charged particle in a magnetic field in the positive $\hat{z}$-direction, i.e. it couples to the cyclotron motion of an electron, hence it is also referred to as the cyclotron resonance active (CRA) wave. The right-circularly polarized (+) mode is consequently called the cyclotron resonance inactive (CRI) wave. By inserting the Drude conductivity into Eq. \ref{eq3}, we can express $\epsilon(B,\omega)$ as 

\begin{equation}\label{eq4}
    \begin{split}
    n_{\pm}^{2}&=\epsilon_{\pm}(B,\omega)=\\
    &\epsilon_{r}\left(1-\frac{\omega_{p}^{2}}{\omega(\omega\pm\omega_{c}+i\nu)}+\frac{\omega_{L}^{2}-\omega_{T}^{2}}{\omega_{T}^{2}-\omega^{2}-i\gamma_{ph}\omega}\right),
    \end{split}
\end{equation}

\noindent where $\omega_{p}=\sqrt{\frac{ne^{2}}{m^{*}\epsilon_{r}\epsilon_{0}}}$ is the plasma frequency, $\omega_{c}=eB/m^{*}$ is the cyclotron resonance frequency and $\nu$ is the scattering rate. To include the effect of a non-parabolic conduction band, we have used $\mathbf{k\cdot p}$ perturbation theory to self-consistently calculated the Landau fan diagram, giving us the magnetic field dependence of the effective mass \cite{PhysRev.115.1172,PhysRev.122.31}. Based on the observation that the sample already enters the quantum limit at 1.3 T (see Fig. \ref{fig:1}), we calculate the magneto-transmission spectra using the cyclotron frequency $\omega_{c}$ corresponding to the transition between the first and second Landau level with spin-up electrons, assuming that $m^{*}(B=0)=0.014m_{e}$, $E_{g}=0.24$ eV and $\Delta_{0}=0.82$ eV. It should be noted that the transmitted light will be a superposition of both propagation modes, which means that the transmittance is given by 

\begin{equation}\label{eq5}
    T=\left|\frac{\tau_{+}+\tau_{-}}{2}\right|^{2}=\frac{|\tau_{+}|^{2}+|\tau_{-}|^{2}+2\mathrm{Re}(\tau_{+}\tau_{-}^{*})}{4},
\end{equation}

\noindent where $\tau_{\pm}$ are the transmission coefficients of the relevant propagation mode for a plane parallel slab as in Ref.~\onlinecite{Palik_1970}. The last term on the right hand side represents the interference between the two circularly polarized propagation modes and is proportional to the phase difference. The magneto-transmission spectra depicted in Fig. \ref{fig:2}c and d are obtain with Eq. \ref{eq4} and \ref{eq5} by setting $n=1.41\times10^{15}$ cm$^{-3}$ and $\nu$=1 THz. A direct comparison between the simulation and the experimental data reveals that most features are indeed reproduced by this simple model. It is worth mentioning that the quality of the simulation strongly depends on the scattering rate as it not only affects the shape of the resonance, i.e. width and depth of the dip in transmittance, but also the amplitude of interference fringes. The value of 1 THz, which is in reasonable agreement with the scattering rate extracted from the electron mobility ($\nu=1.51$ THz), was chosen as it gives the best match between theory and measurement. Since two-phonon processes are not accounted for within the framework of the harmonic oscillator approximation, the opacity for radiation with frequencies larger than 15 THz, related to the 2LO phonon mode, is not captured.

\subsection{FIR-induced thermoelectric signal}

\begin{figure}[t]
    \centering
    \includegraphics[width=8.4cm]{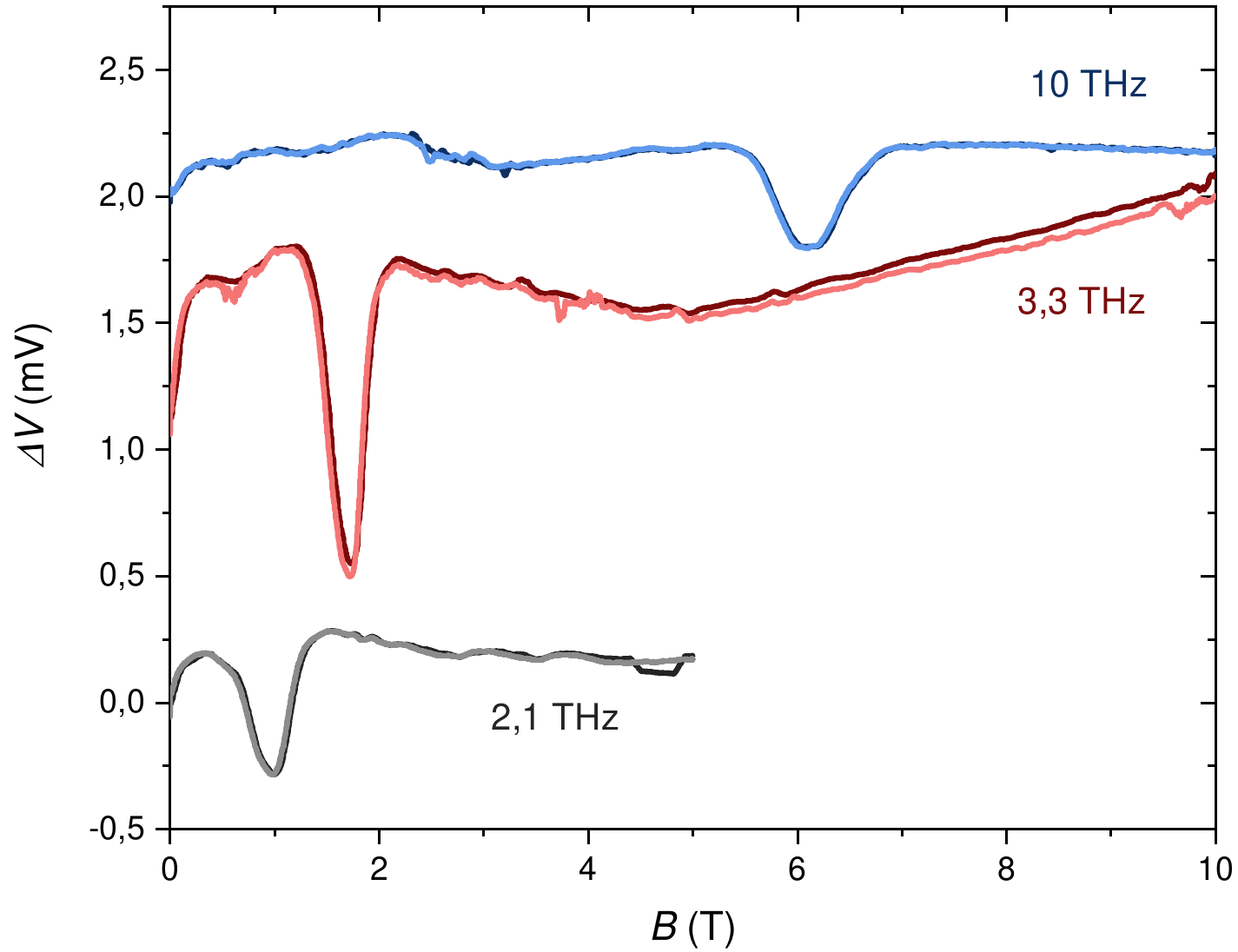}
    \caption{Light-induced thermoelectric voltage as a function of magnetic field for different frequencies in a slab of weakly doped InSb at $T=4.2$ K. The measurements at each frequency were performed at a positive bias of 10 mA (dark trace) and a negative bias of -10 mA (light trace) without attenuation. For all frequencies the incident power was of the order of 10 mW.}
    \label{fig:3}
\end{figure}

\begin{figure*}[t]
    \centering
    \includegraphics[width=16cm]{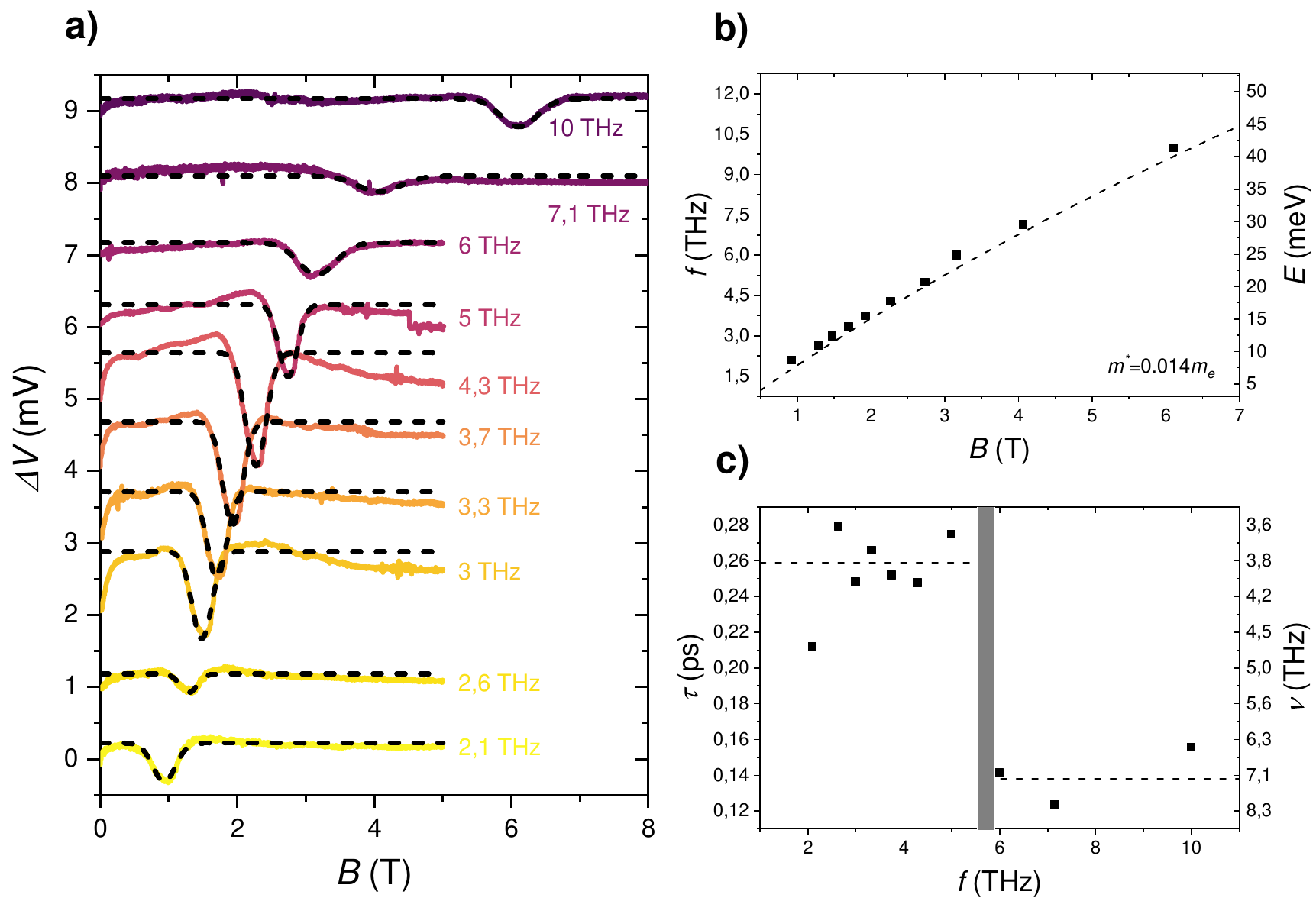}
    \caption{(a) Light-induced thermoelectric voltage as a function of magnetic field for different frequencies in a slab of weakly doped InSb at $T=4.2$ K. The sample was biased with a current of 10 mA. The traces have a constant offset of 1 mV. Generally, the incident power was of the order of 10 mW. The position of the cyclotron resonance, i.e. the dip in the thermoelectric signal, is determined with a Gaussian fit (black dashed line). (b) Magnetic field dependence of the cyclotron energy. The black dashed line indicates the theoretically predicted position of the cyclotron resonance based on $\mathbf{k\cdot p}$ theory. (c) Frequency dependence of the scattering rate. The shaded region shows the position of the TO and LO phonon band.}
    \label{fig:4}
\end{figure*}

In the upcoming section, we will discuss light-induced voltages that were acquired by exposing the sample to high fluences, i.e. using a waveguide to couple the light into our setup. Figure \ref{fig:3} shows the magnetic field dependence of light-induced voltages for both bias polarities at different frequencies in a slab of nominally undoped InSb. As these light-induced voltages seem to be independent of bias, we can state that the resistive contribution to the signal is negligible. Moreover, the signal even persists when the sample is unbiased. We therefore propose that these light-induced voltages have a thermoelectric origin, stemming from the thermal gradient created by locally heating the sample with the laser. If we compare our light-induced voltages to thermopower experiments on InSb, which are presented in the doctoral thesis of M. Shahrokhvand, we can conclude that the thermoelectric response is exceptionally large \cite{shahrokhvand2019magneto}. This is especially true for small magnetic fields where the Seebeck coefficient $S_{xx}$ is of the order of 1 $\mu$V/K and the light-induced voltages would correspond to a temperature increase of tenths of Kelvin. However, a direct comparison between our data and thermopower experiments in not possible, as the Seebeck coefficient was determined using the assumption that $S\propto\frac{\Delta V}{\Delta T}$. For our measurements, we are not in this linear regime and have to take into account higher order contributions, since the electronic system is strongly heated by the light. Still, the thermoelectric response is unexpectedly large, considering that the mask ensures that only the center of the sample is illuminated, creating a radially symmetric temperature gradient where one would not expect a net photocurrent to appear due to the symmetry of the exposure. However, if the Gaussian beam profile of the laser spot is distorted or even partly cut off by the mask, which cannot be ruled out because of the limited control over the optical path when using the waveguide, the symmetry is broken and one would expect a thermoelectric contribution in the electric signal. 

Yet, we can extract some useful information from this purely thermoelectric response to IR radiation. The thermoelectric response of the sample exhibits a pronounced dip at certain magnetic fields depending on the frequency. To investigate the dispersion of this resonance-like feature, we have performed these measurements at several frequencies (see Fig. \ref{fig:4}a). By plotting these frequencies against the magnetic field positions and comparing the resulting dispersion to the theoretically predicted position of the cyclotron resonance, we notice the excellent agreement between data and theory (see Fig. \ref{fig:4}b). The theoretical position of the cyclotron resonance is again determined from the Landau fan diagram constructed with $\mathbf{k\cdot p}$ perturbation theory. Based on the dispersion, similar to the magneto-transmission spectra, the cyclotron resonances are governed by transitions between the first and second Landau level with spin-up electrons with an effective mass of $m^{*}(B=0)=0.014m_{e}$. From the Gaussian fit of these cyclotron resonance features we extract the lifetimes $\tau=\frac{\hbar}{\Gamma}$ from the FWHM, where $\Gamma=\frac{\hbar e\Delta B}{m^{*}}$ is the width of the resonance expressed in units of energy. The evolution of the lifetime as function of frequency is presented in Fig. \ref{fig:4}c. In this figure, we observe an abrupt drop in lifetime after passing through the Reststrahlen band. Such an enhancement of the scattering rate has been seen before in transmission experiments as a broadening of the cyclotron resonance, where the electron-phonon coupling renormalizes the Landau levels and opens up additional scattering channels \cite{VIGNERON1979595,PhysRevB.57.12156}. However, one should realize that the corresponding scattering rates are several times larger than those derived from the mobility or magneto-transmission spectra. A possible explanation for this discrepancy could be related to the thermoelectric origin of the signal, which means that the relaxation mechanisms will be different as we not only probe the electronic system.

\begin{figure}[t]
    \centering
    \includegraphics[width=8.4cm]{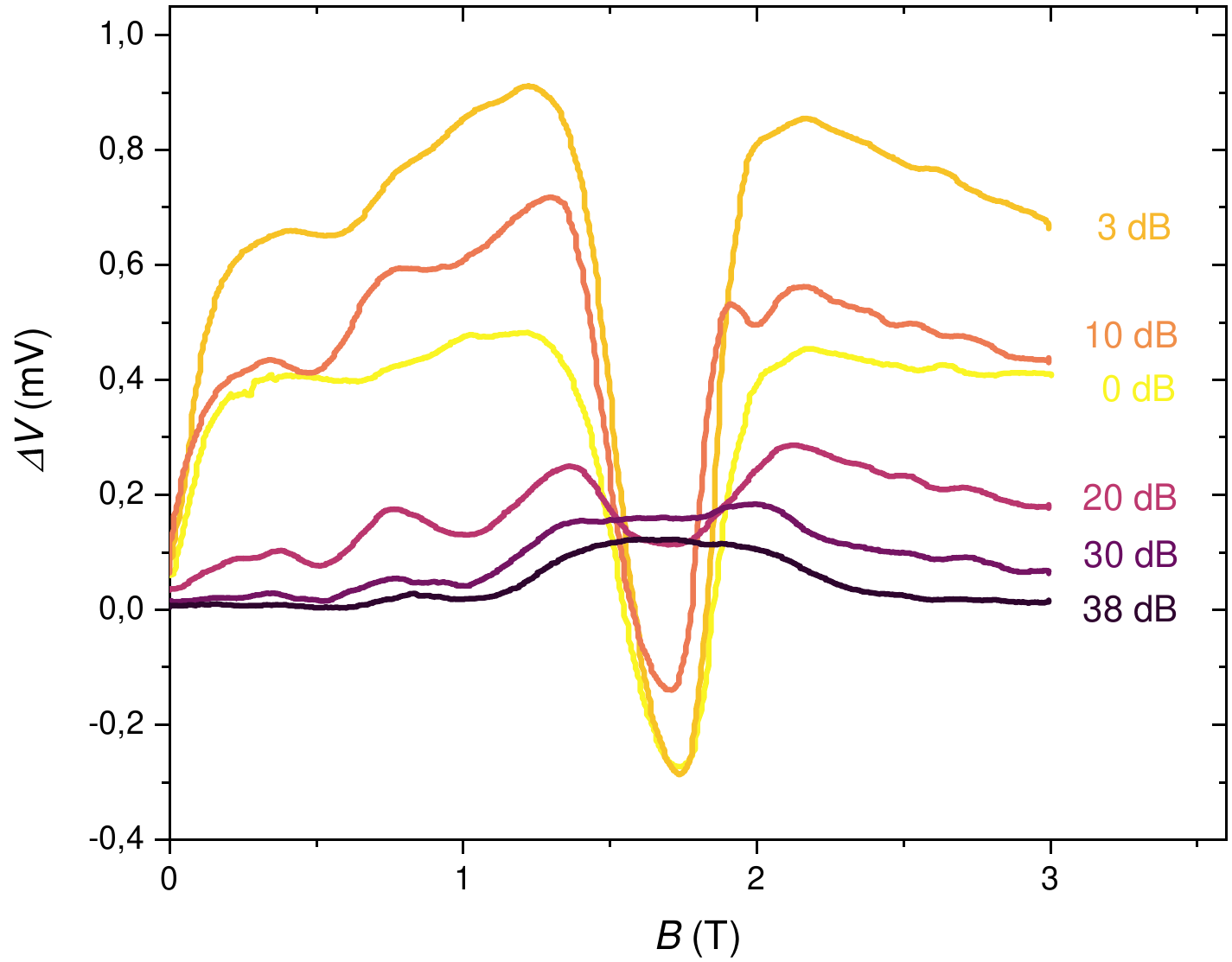}
    \caption{Light-induced thermoelectric voltage as a function of magnetic field at $f$=3.3 THz for different levels of power attenuation ($P_{0 dB}\approx 10$ mW). The measurements were performed at $T=4.2$ K.}
    \label{fig:5}
\end{figure}

Looking at the power dependence of these light-induced voltages in Fig. \ref{fig:5}, we notice that there is a non-monotonic relation between the signal amplitude and the fluence of the laser. This trend is somewhat counterintuitive, as a larger fluence produces a bigger temperature gradient, which in turn should result in a stronger thermoelectric response. A possible cause for this unexpected behaviour can be found, by inspecting the Mott relation for diffusive thermoelectric power (TEP) $S_{d}$. The magneto-transmission spectra have revealed that the conductivity of the material is still governed by the Drude expression, which allows us to rewrite the Mott relation as follows

\begin{equation}\label{eq6}
    S_{d}=\frac{\pi^{2}k_{B}^{2}T}{3eE_{F}}(3+p),
\end{equation}

\noindent where $p=\frac{\partial\mathrm{ln}\tau}{\partial\mathrm{ln}E}\big|_{E_{F}}$ \cite{PhysRevLett.102.096807,okazaki2012photo}. Thus, the non-monotonic power dependence can be understood by realizing that at the largest fluences an increase in carrier concentration, brought about by the thermal activation of carriers, leads to a larger Fermi energy $E_{F}$ and reduces the thermopower of the sample (see Eq. \ref{eq6}). In Fig. \ref{fig:5}, we also see that the cyclotron resonance in the thermoelectric response at some point changes sign as the power is reduced, suggesting that the induced photocurrent reverses its direction. In general the beam profile does not change significantly when using attenuators, i.e. the orientation of the temperature gradient will remain unchanged, and can consequently not be the origin of this sign reversal. A closer inspection of the trace for 30 dB of attenuation, reveals that there are possibly two contributions to the signal, as we can make out a shallow dip which is superimposed on top of a broad peak. Furthermore, these two contributions have different relaxation mechanisms as the resonance becomes gradually broader when decreasing the fluence. It is worth pointing out that at the highest fluence the width of the resonance becomes larger again, which supports our previous conclusion that we raise the average sample temperature with the incident IR radiation.

\begin{figure*}[t]
    \centering
    \includegraphics[width=16cm]{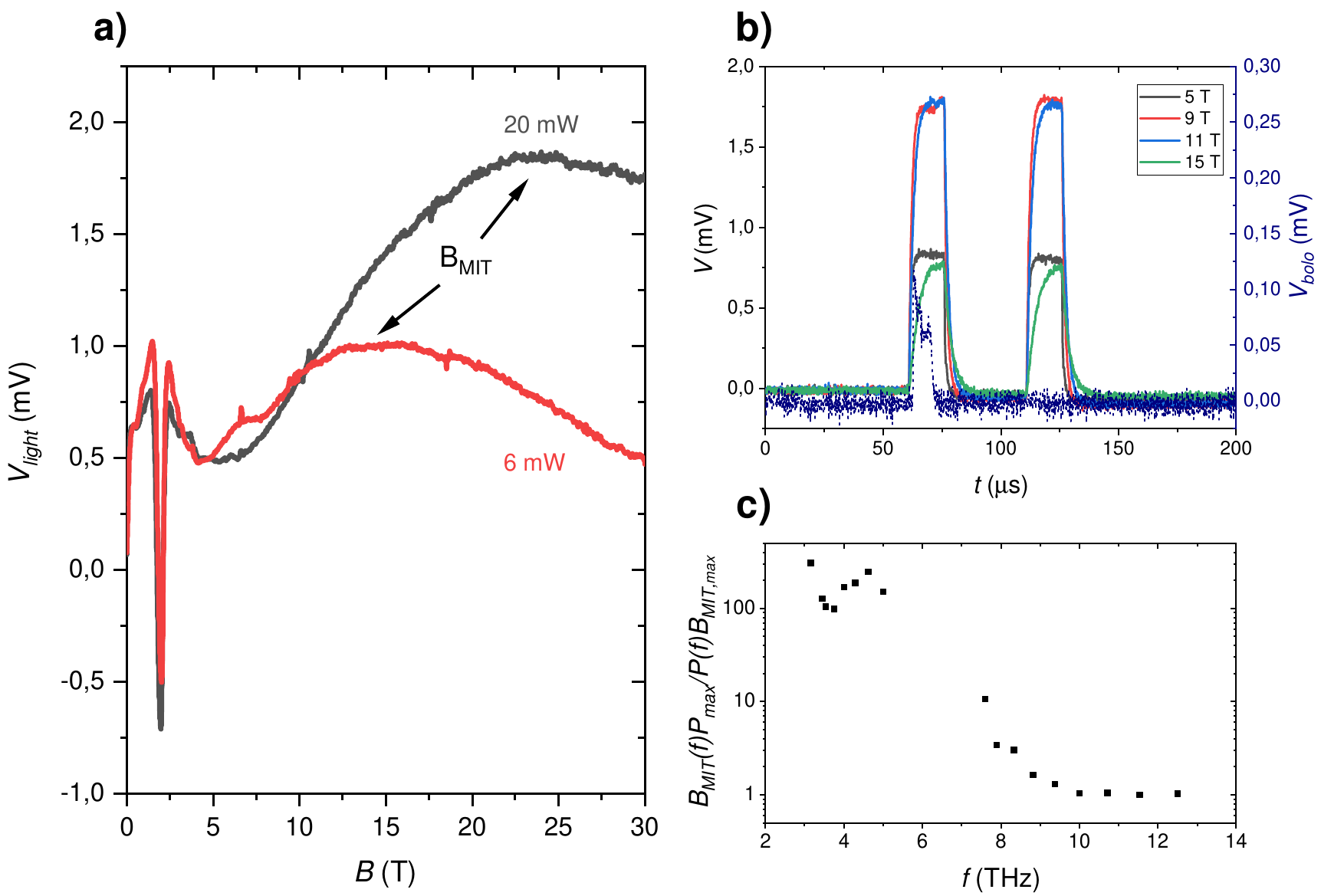}
    \caption{(a) Light-induced thermoelectric voltage as a function of magnetic field for $f$=3.7 THz at two different light intensities in a slab of weakly doped InSb. The sample was biased with a current of 10 mA. (b) Averaged traces recorded with the oscilloscope at different magnetic fields for $f$=12.5 THz. The response of the bolometer is included as a reference to illustrate the effect of the light pulse (dashed line). (c) Frequency dependence of the normalized inflection point of the light-induced thermoelectric voltage. All measurements were performed at at $T=4.2$ K.}
    \label{fig:6}
\end{figure*}

Let us now take a step back and evaluate the on-light signal, i.e. the light-induced voltages without subtraction of the reference signal. In this `bare' thermoelectric response, displayed in Fig. \ref{fig:6}a, we not only notice cyclotron resonances, but also the presence of maxima at high magnetic fields. Such a maximum is also observed in thermopower measurements on InSb and is attributed to the metal-insulator transition (MIT) \cite{shahrokhvand2019magneto}. The onset of the MIT can be predicted by the magnetic freeze-out condition, which is based on the Mott criterion, as in Ref.~\onlinecite{Zeitler_1994}

\begin{equation}\label{eq7}
    (a_{\parallel}a_{\perp}^{2}n)^{\frac{1}{3}}\approx0.3,
\end{equation}

\noindent where $a_{\parallel}=\frac{a_{B}^{*}}{\ln{(a_{B}^{*}/l_{B})}^{2}}$ and $a_{\perp}=2l_{B}$ are the dimensions of the wave function parallel and perpendicular to the magnetic field, respectively. $a_{B}^{*}$ is the effective Bohr radius and $l_{B}=\sqrt{\frac{\hbar}{eB}}$ is the magnetic length. Using Eq. \ref{eq7}, we expect the MIT to occur at $B_{MIT}=3.8$ T, which disagrees with the value one can extract from the thermoelectric response of the sample. However, the transition depends on the fluence of the laser, suggesting that the carrier concentration is modulated by locally heating the sample. The maxima in Fig. \ref{fig:6}a would correspond to a carrier concentration of $n=14\times10^{15}$ cm$^{-3}$ and $n=8\times10^{15}$ cm$^{-3}$ for the high and low power trace, respectively. Such values of the carrier concentration are unrealistic, as the Hall data points towards an increase of only 3\% when raising the temperature to 160 K. In the raw traces obtained with the oscilloscope, we notice that the waveform of the current pulses become distorted after passing through the `MIT' in the thermoelectric response (see Fig. \ref{fig:6}b). We attribute this distortion to the circuit time constant getting larger when the sample enters the insulating state. The feature is thus linked to the MIT, but does not match the exact onset of this transition, since it is most likely governed by the resistance of the sample. Apart from the apparent effect that fluence has on the MIT, we also checked for a possible frequency dependence. For this purpose we will introduce the following figure of merit $\frac{B_{MIT}(f)}{P(f)}\frac{P_{max}}{B_{MIT,max}}$, where $P_{max}$ and $B_{MIT,max}$ are the largest values of the individual quantities in our data set. This metric accounts for the fact that the power varies for different frequencies and with it the position of the MIT. The frequency dependence of this metric in Fig. \ref{fig:6}c points towards an abrupt change in the light-matter interaction, since the MIT becomes less sensitive to the fluence of the laser for frequencies larger than 6 THz, which roughly corresponds to the position of the Reststrahlen band. This observation supports our previous statement that additional scattering channels are opened up as the frequency exceeds that of the phonon modes, suggesting that the sample can dissipate the energy of the light more efficiently and hence produce less thermally activated carriers.

\begin{figure*}
    \centering
    \includegraphics[width=16cm]{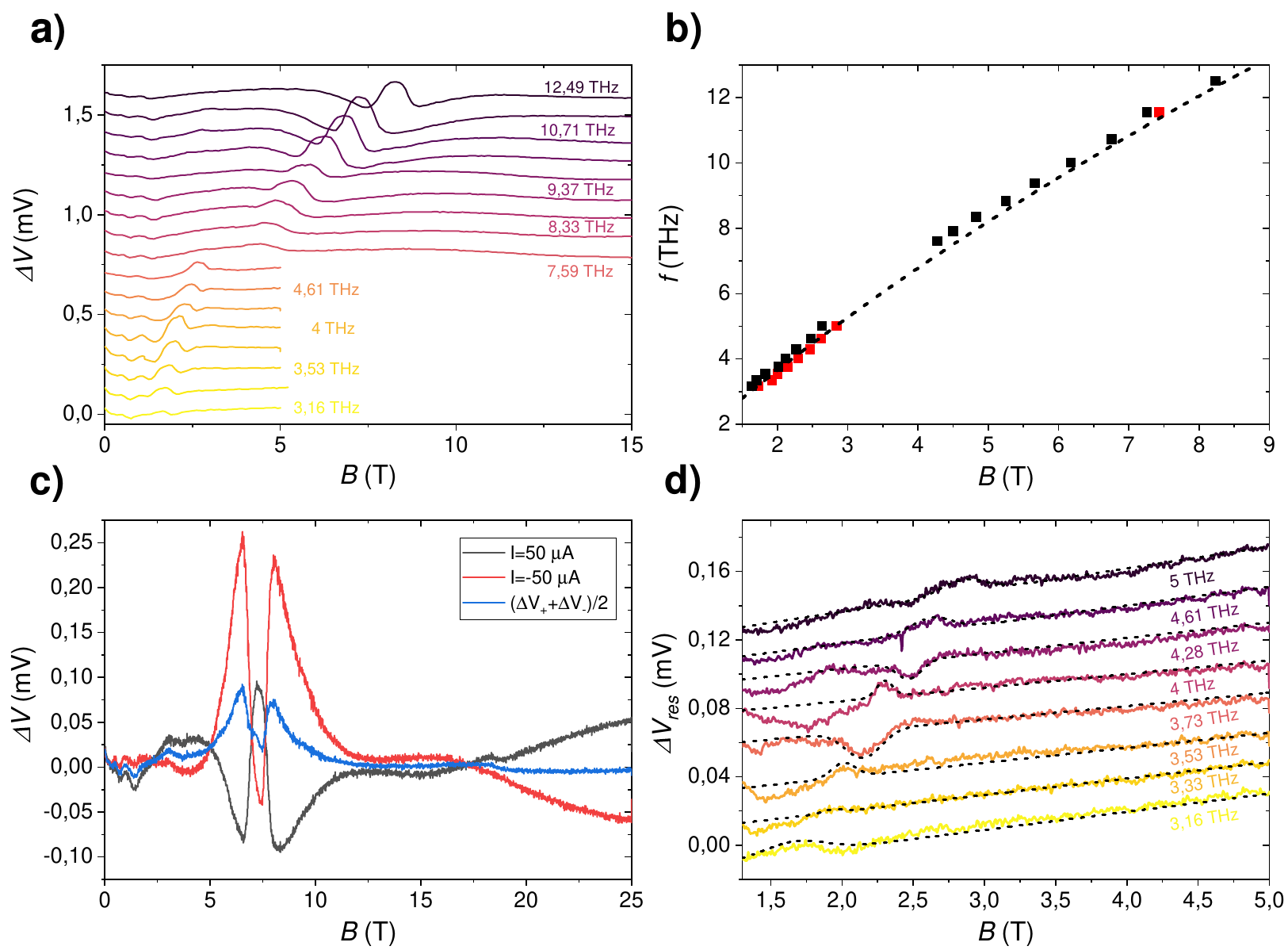}
    \caption{a) Magnetic field dependence of light-induced voltages that were measured at different frequencies in a slab of weakly doped InSb . The sample was biased with 50 $\mu$A current pulses. The curves have a constant offset of 0.1 mV. (b) Magnetic field dependence of the cyclotron resonance feature in the electric (black) and purely resistive signal (red). The black dashed line indicates the theoretically predicted position of the cyclotron resonance based on $\mathbf{k\cdot p}$ theory. (c) shows the light-induced voltages for positive and negative biasing, as well as voltages related to resistance changes $\frac{\Delta V_{+}-\Delta V_{-}}{2}$ for $f=11.53$ THz. (d) Magnetic field dependence of voltages stemming from light-induced resistance changes that were measured at different frequencies in a slab of nominally undoped InSb. The curves have a constant offset of 0.02 mV. The black dashed line, consisting of a Lorentzian with a linear background, acts as a guide to the eye for highlighting the position of the cyclotron resonance. All measurements were performed at $T=4.2$ K.}
    \label{fig:7}
\end{figure*}

\subsection{Resistively detected cyclotron resonance}
The light-induced response in the electric signal discussed so far had a purely themoelectric origin. To investigate the possibility of resistively detecting cyclotron resonances, we have used the free space optics setup to attain more control over how the laser spot hits the sample. The use of the free space optics should reduce the thermoelectric contribution and increase our chances of identifying features related to resistively detected cyclotron resonances. Figure \ref{fig:7}a shows the magnetic field dependence of the light-induced voltages for different frequencies. Here, we observe a peak at certain magnetic fields depending on the frequency of the light, which is comparable to the cyclotron resonance feature in Fig. \ref{fig:5} for the lowest fluence. The dispersion of the cyclotron resonance plotted in Fig. \ref{fig:7}b (black squares), where the peak positions are determined by fitting the data with a Gaussian, is governed by the same LL transitions as in the magneto-transmission spectra.

The similarity between the light-induced voltages in Fig. \ref{fig:7}a and the trace for the smallest power in Fig. \ref{fig:5}, still points towards the presence of a pronounced thermoelectric contribution. In order to determine the light-induced voltage that stems from a resistance change, we acquire the photoconductive response of the sample for positive and negative current pulses and define the resistive signal as $\Delta V_{res}=\frac{\Delta V_{+}-\Delta V_{-}}{2}$, thereby excluding any thermoeltric contributions (see Fig. \ref{fig:7}c). Figure \ref{fig:7}d displays $\Delta V_{res}$ as a function of magnetic field at different frequencies. In these traces, we can again discern resonance-like features which nicely coincide with the theoretically predicted position of the cyclotron resonance (see red squares in Fig. \ref{fig:7}b). Although, the sign of the resonance is still peculiar, considering the sample has already entered the quantum limit, meaning that the resistance should only drop when the system heats up. Looking at the signal strength of the resistively detected cyclotron resonance, which can be up to two order of magnitudes smaller than the resonance in the light-induced voltages, it might be more convenient to use the thermoelectric response instead. The possibility of utilizing the thermoelectric response of the system to detect resonances was already demonstrated by Kinoshita \textit{et al.}, who used the photo-Nernst effect to measure CRs in graphene devices \cite{Kinoshita}.

\section{Conclusion}
To conclude, by performing FIR transmission experiments we have discovered that the electronic properties for frequencies far beyond the scattering rate of nominally undoped InSb are fully governed by the Drude conductivity, resulting in an excellent agreement between data and simulation. Additionally, we have presented initial proof-of-principle experiments for resistively detecting CRs and have discovered a large thermoelectric contribution in the photoconductive response. The thermoelectric signal contains a prominent resonance-like feature, which moves to higher magnetic fields when increasing the frequency. The corresponding magnetic field dependence of this feature matches nicely with the predicted position of the CR obtained from $\mathbf{k\cdot p}$ perturbation theory, proving that the feature is indeed linked to a cyclotron resonance. From the power dependence of this feature we conclude that the thermoelectric signal consists of multiple contributions with different directionality and relaxation mechanisms. In spite of the rather large thermoelectric signal, we are still able to observe resonant heating at CR conditions after isolating the weak resistive signal. But there are still some aspects of the resistive signal that are not fully understood, e.g. the sign of the modulation changes with frequency.

\begin{acknowledgments}
This work was supported by HFML-FELIX, member of the European Magnetic Field Laboratory (EMFL). It is part of the research programme “HFML-FELIX: a unique research infrastructure in the Netherlands. Matter under extreme conditions of intense infrared radiation and high magnetic fields” with project number 184.034.022 financed by the Dutch Research Council (NWO).
\end{acknowledgments}

\bibliography{refs}

\end{document}

%% file: preamble.tex
\usepackage{amsthm}
\usepackage{mathtools}
\usepackage{physics}
\usepackage{xcolor}
\usepackage{graphicx}
\usepackage[left=23mm,right=13mm,top=35mm,columnsep=15pt]{geometry} 
\usepackage{adjustbox}
\usepackage{placeins}
\usepackage[T1]{fontenc}
\usepackage{lipsum}
\usepackage{csquotes}
\usepackage{floatrow}
\usepackage{float}

\usepackage{nicefrac}
\usepackage{amsfonts}

\usepackage{amssymb}
\usepackage{amsmath} 
\usepackage[caption=false]{subfig}
\usepackage{multirow} 
\usepackage{tabularx} 
\usepackage{array}

\usepackage{units}

\usepackage{tensor} 
\usepackage{braket}

\usepackage{bm}
\usepackage{hyperref}

